\documentclass[twocolumn,prl,aps,floatfix,superscriptaddress,amssymb]{revtex4}
\usepackage{epsfig}
\usepackage{amsmath}

\begin{document}

\title{Electric-field controlled spin in bilayer triangular graphene quantum dots}

\author{A.~D.~G\"u\c{c}l\"u}
\affiliation{Institute for Microstructural Sciences, National Research Council of Canada
, Ottawa, Canada}

\author{P.~Potasz}
\affiliation{Institute for Microstructural Sciences, National Research Council of Canada
, Ottawa, Canada}
\affiliation{Institute of Physics, Wroclaw University of Technology, Wroclaw, Poland}

\author{P.~Hawrylak}
\affiliation{Institute for Microstructural Sciences, National Research Council of Canada
, Ottawa, Canada}

\date{\today}

\begin{abstract}
We present theoretical results based on mean-field and exact many-body
approaches showing that in bilayer triangular graphene quantum dots with
zigzag edges the magnetism can be controlled by an external vertical
electric-field. We demonstrate that without electric field the spins of the
two layers of the quantum dot interact ferromagnetically. At a critical value
of the electric-field, the total spin of the bilayer structure can be turned
off or reduced to  a single localized  spin, a qubit isolated from contacts
and free from interaction with nuclear spins.
\end{abstract}

\maketitle

Graphene exhibits unusual electronic
properties\cite{Wallace,NGM+04,NGM+05,ZTS+05,ZGG+06,NGP+09} including
relativistic nature of quasi-particles, sublattice pseudospin, and zero energy
bandgap. The application of graphene for logic devices requires opening of the
bandgap which can be achieved by either size quantization
\cite{PSK+08,LYG+11,GFS+10,WSG08,AHM08,Eza10,GPV+09,GPH10,PGH10,VGP+11,FP07,WMK08},
or chemical modification\cite{ENM+09}, or bringing in a second layer and
applying an external electric-field
\cite{MAF07,OBS+06,CNM+07,OHL+07,MLS+09,ZTG+09,WAF+10,CNM+10,ZLB+08}, or size
quantization in bilayer graphene
nanostructures\cite{PVP07,CPS+08,SMM+08,WCZ09}. In particular, when graphene
is reduced to a triangular quantum dot with zigzag edges\cite{LYG+11},
sublattice symmetry is broken, size dependent energy gap is open, and a band
of degenerate states at the Fermi level is created leading to finite
macroscopic spin polarization
\cite{AHM08,Eza10,GPV+09,GPH10,PGH10,VGP+11,FP07,WMK08}. This allows
simultaneous size, shape and edge engineering of magnetic, electrical, and
optical properties in a single material-graphene\cite{GPH10}. 

In this work, we investigate electronic and magnetic properties of bilayer
triangular graphene quantum dots with zigzag edges under external vertical
electric-field. We show that the magnetic moment of bilayer triangular
graphene quantum dots can be controlled by the vertical
electric-field. Without the electric-field, the magnetic moments of the two
layers are shown to be coupled ferromagnetically. Using configuration
interaction and mean-field calculations based on tight-binding model, we
demonstrate that the ferromagnetism can be either turned off or reduced to a
single electron/hole spin. The single electron spin is hence isolated in a
charge neutral structure by the application of an electric-field, independent
of the size of the quantum dot and without decoherence due to contacts. The
electric-field control of the ferromagnetism \cite{OCM+00} and isolation of a
single spin opens new applications in spintronics and quantum information
processing\cite{RTB07,SCL06,TBL+07,SSC10,AKK10}. 

\begin{figure}
\epsfig{file=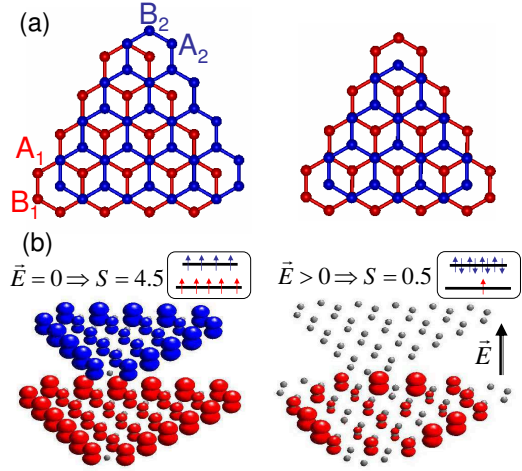,width=2.8in}
\caption{\label{fig:1} (Color online)   Bilayer triangular graphene quantum
  dot with zigzag edges. (a) structure constructed using two single layer
  quantum dots with equal sizes (on the left) and different sizes (on the
  right). (b) Isosurface plot of the spin density
  $\rho_\uparrow-\rho_\downarrow$ for with and without electric-field ${\bf
    E}$, obtained from configuration interaction calculations.}
\end{figure}

Figure 1a shows two possibilities for building a bilayer triangular graphene
quantum dot (BQD) using two single layer triangular quantum dots (TGQD) of
comparable sizes, with zigzag edges. We consider AB Bernal stacking, where the
A sublattice of the top layer (A2, shown in blue color) is on top of the B
sublattice of the bottom layer (B1, shown in red). On the left hand side, the
two TGQDs are of the same size. In this configuration, however, not all the A2
atoms have a B1 partner as required by Bernal stacking. A more natural
configuration choice is shown on the right hand side of Fig.1a. The top layer
triangle has its floating atoms removed, making it smaller than the bottom
layer triangle. Such a bilayer
construction has the interesting property of having an odd number of
degenerate  states at the Fermi level (zero-energy shell of states)
independent of its size, allowing to isolate a single spin in a charge neutral
structure and hence isolated from the contacts. This is illustrated in Fig.1b,
where the spin density isosurfaces are shown for zero electric-field (left
hand side) and finite electric-field (right hand side), as obtained from our
tight-binding based configuration interaction calculations explained in detail
below. When the electric-field is off, both layers have a finite magnetic moment,
as in single layer triangles\cite{Eza10,GPV+09,GPH10,PGH10,VGP+11,FP07,WMK08},
differing by one spin due to the size difference of the two triangles. The
magnetic moments of the two layers are coupled ferromagnetically. When a
sufficiently high electric field is applied, electrons from the lower layer
reduce their energy by  flipping their spin and transfering to the top layer
filling up all the available spin up and down zero-energy states, leaving
behind one single spin.

\begin{figure}
\epsfig{file=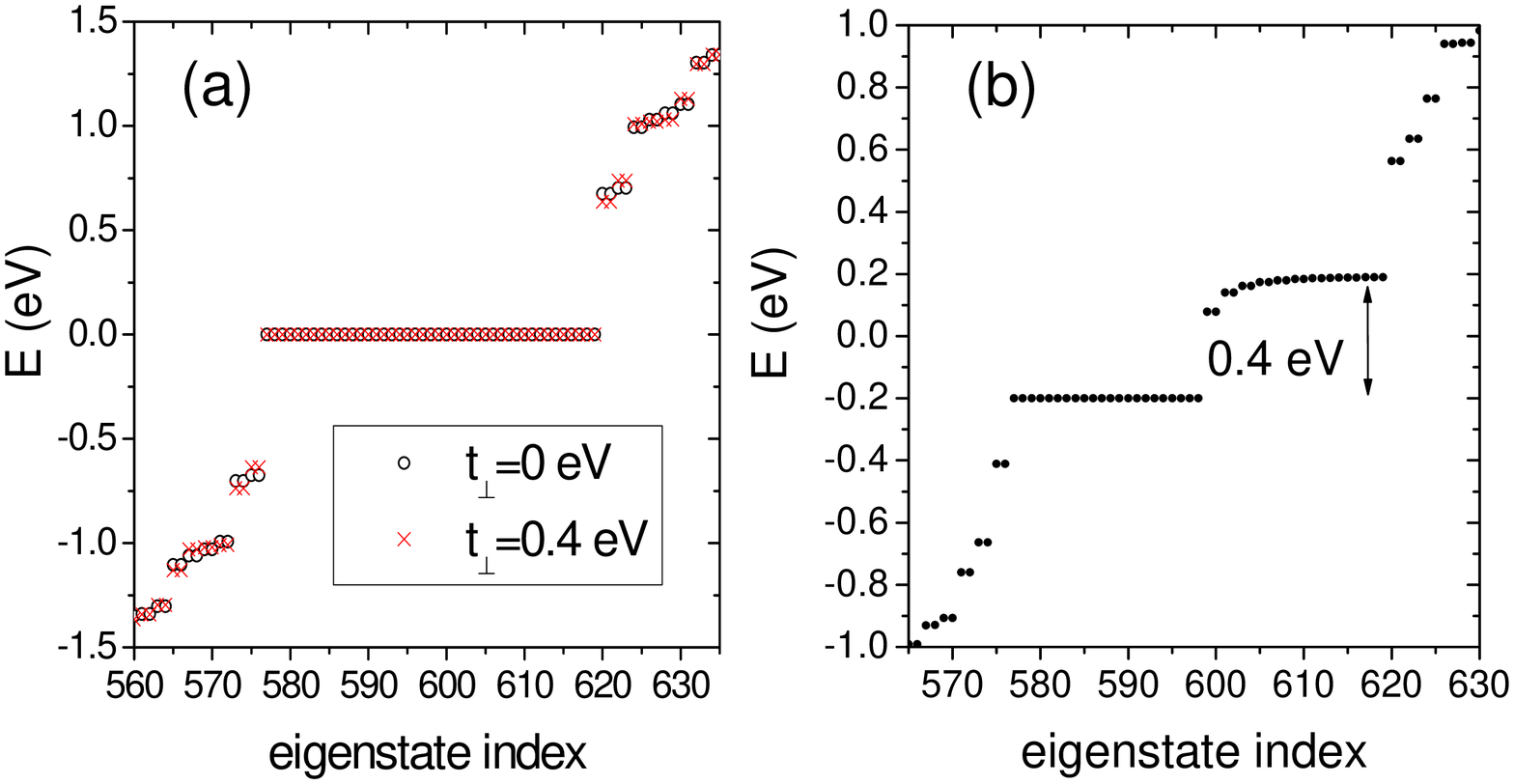,width=2.8in}
\caption{\label{fig:2} (Color online)  
Single particle tight-binding spectrum. (a) The bilayer quantum dot consisting
of 1195 atoms has 43 zero-energy states. (b) When an electric-field is applied, the
degeneracy between the 21 top layer zero-energy states and 22 bottom layer
zero-energy states is lifted.}
\end{figure}

We have confirmed the above picture through calculations with varying levels
of accuracy. As a first step, we diagonalize the tight-binding Hamiltonian
given by 
\begin{eqnarray}
H_{TB}= \sum_{ij\sigma}\tau_{ij}c^\dagger_{i\sigma}c_{j\sigma}
      + \sum_{i\sigma}V_{i}c^\dagger_{i\sigma}c_{i\sigma}
\label{eq:Htb}
\end{eqnarray}
where the operator $c^\dagger_{i\sigma}$ creates an electron on a $p_z$
orbital on site “$i$” with spin $\sigma$. The tight-binding parameters
$\tau_{ij}$ are fixed to their bulk values $t=-2.8$ eV for in-plane nearest
neighbours “$i$” and “$j$” and $t_{\bot}=0.4$ eV for inter-layer hopping. The
effect of the potential difference induced by an external perpendicular
electric-field ${\bf E}$ is taken into account through $V_i=-\Delta V/2$ for
the bottom layer atoms and $V_i=\Delta V/2$ for the top layer atoms. Figure 2a
shows the energy spectrum near the Fermi level for $\Delta V=0$ for a BQD
consisting of 622 atoms in the bottom and 573 atoms in the top layer. If we
take $t_{\bot}=0$, the two triangles are decoupled and we find 22 zero-energy
states in the bottom layer and 21 zero-energy states in the top layer, for a
total of 43 zero-energy states, consistent with previous work on single layer
TGQDs\cite{Eza10,GPV+09,GPH10,PGH10,VGP+11,FP07,WMK08}. Turning on $t_{\bot}$
to $0.4$ eV does not affect the zero-energy states. The effect of applying an
electric field, {\it e.g.}  $\Delta V=0.4$ eV, is shown in Fig.2b. The energy
of the 21 zero-energy states corresponding to the top layer is pushed up by
0.4 eV with respect to the bottom layer zero-energy states. Note that the
bottom layer zero-energy states do not experience any dispersion unlike the
top layer zero-energy states. This is due to the fact that they lie strictly
on A1 sites which are not coupled to the top layer, whereas the top layer
zero-energy states, which lie on B2, do couple to the bottom layer. The
ability of controlling the relative position of zero-energy states gives an
interesting opportunity to control the charge and spin of the zero-energy
states, expected to be spin polarized for $\Delta V=0$ according to Lieb's
theorem\cite{Lieb89}.

\begin{figure}
\epsfig{file=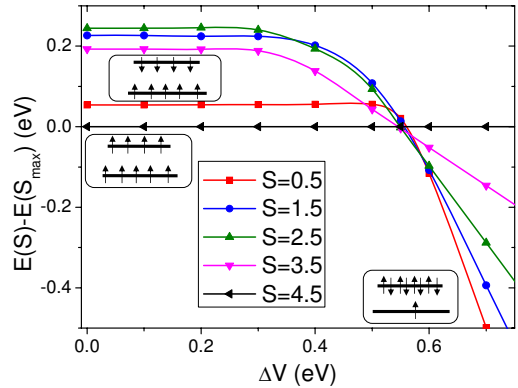,width=2.8in}
\caption{\label{fig:3} (Color online)  
Meanfield Hubbard results. Energies of lowest energy states with different
total spin S as a function of potential difference $\Delta V_c$ between the layers,
with respect to the ferromagnetic configuration $S_{max}=4.5$ for a system of 107
atoms and 9 zero-energy states.}
\end{figure}

The magnetic properties of BQD structures as a function of applied electric
field can be studied by solving the Hubbard model within the self-consistent
mean-field approach:  
\begin{eqnarray}
H_{MF}= \sum_{ij\sigma}\tau_{ij}c^\dagger_{i\sigma}c_{j\sigma}
      + \sum_{i\sigma}V_{i}n_{i\sigma}
      + U \sum_{i\sigma} ( \langle n_{i\sigma} \rangle -\frac{1}{2}) n_{i\sigma}
\label{eq:Hmf}
\end{eqnarray}
where U is the on-site Hubbard term taken to be 2.75 eV here. First, we study
a small BQD of 107 atoms and 9 zero-energy states. Figure 3 shows the energies
for different total spin $S$ with respect to the energy of the ferromagnetic
configuration, $S=9/2$.  At $\Delta V=0$, the degenerate band of zero-energy states is
polarized: all 9 electrons occupying the 9 zero-energy states have their
spins aligned ferromagnetically. Although the first excited state obtained
from the Hubbard model is antiferromagnetic with the total spin of the bottom
layer opposite to the total spin of the top layer, a full treatment of the
correlation effects shows that (see below) low lying excited states have more
complex spin structures. The Hubbard model is, however, useful for estimating
the critical value $V_c$ where  phase transition occurs. As $\Delta V$ is increased,
the electrons lying on the bottom layer zero-energy states are forced to flip
their spin and tunnel to the top layer zero-energy states. At around $\Delta V_c=0.55$
eV such charge transfers occur abruptly, leading to a decrease of the total
spin of the system. As a result, all top layer zero-energy states become
doubly occupied, leaving exactly one single spin in the bottom layer
zero-energy states. We note that one can also isolate a single hole spin in
the bottom layer by applying a reverse electric field, thus pushing the
electrons from the top layer to the bottom layer, occupying all states except
one. It is thus possible to isolate and manipulate a single electron or hole
spin in a neutral BQD by applying an external electric field.

\begin{figure}
\epsfig{file=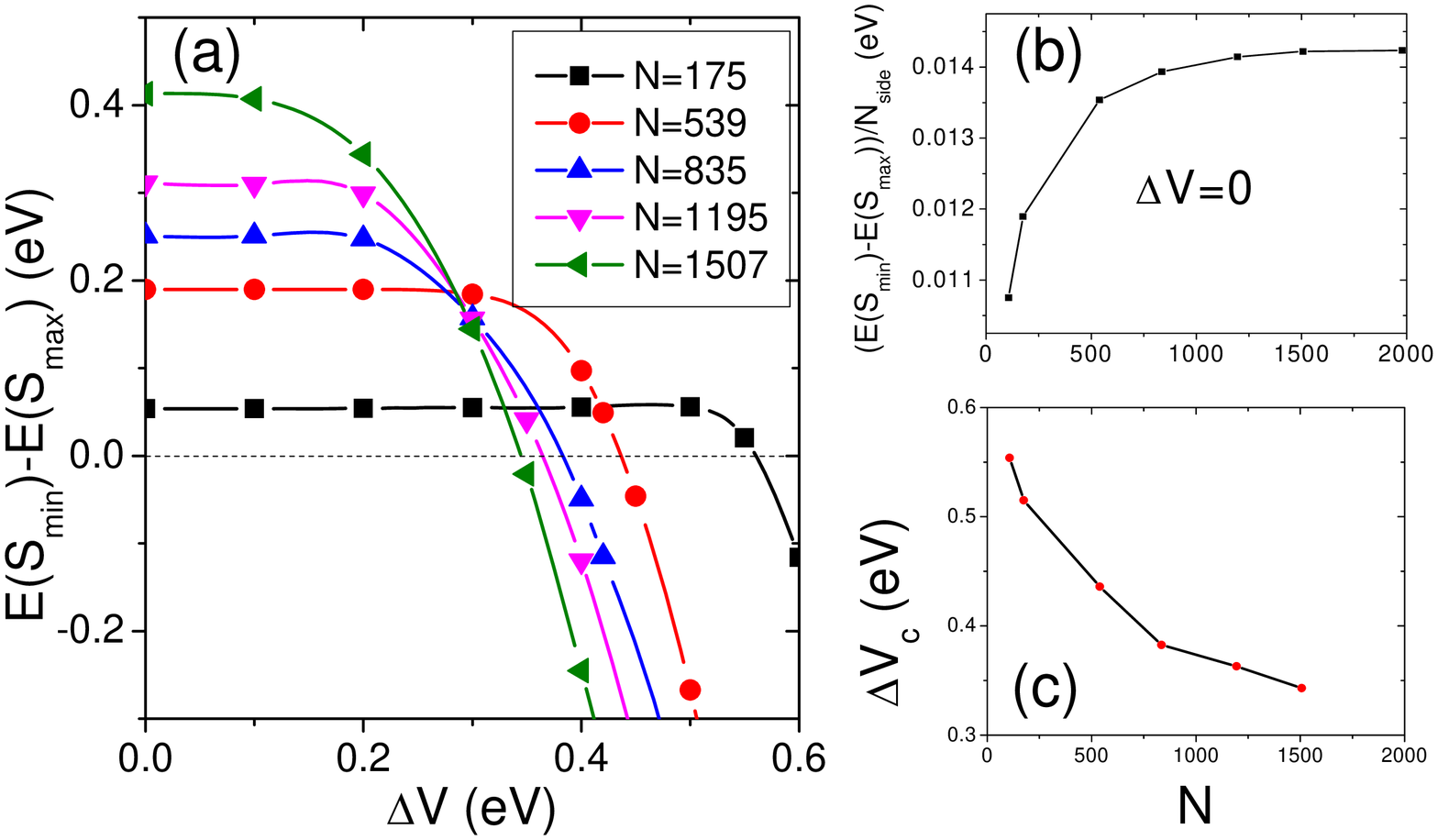,width=2.8in}
\caption{\label{fig:4} (Color online)  
Size dependence of ferromagnetic-antiferromagnetic transition. (a) FM-AFM
energy difference $E(S_{min})-E(S_{max})$ as a function of potential
difference $\Delta V$
between layers, up to $N=1507$ atoms. (b) For $\Delta V=0$, the FM-AFM energy gap per
number of side atoms $N_{side}$ approaches 14.3 meV (top panel). Critical value $\Delta V_c$
where the transition occurs as a function of number of atoms $N$ (bottom
panel). }
\end{figure}

The procedure of isolating single electron or hole should occur regardless of
the size of the system since the top layer has always one less zero-energy
state than the bottom one. In order to investigate the size dependence, in
Fig.4a we show the energy difference between the ferromagnetic and
antiferromagnetic (FM-AFM) states calculated in the mean-field Hubbard
approximation as function of applied voltage for several sizes up to 1507
atoms.  We note that, due to the unusually high degeneracy of the states,
self-consistent iterations occasionally get trapped in a local energy
minima. We have thus repeated the calculations several times using different
initial conditions and/or convergence schemes to assure that the correct
ground state was reached. As expected, at $\Delta V=0$, the FM-AFM gap
increases with the size of the system $N$. In fact, the FM-AFM gap energy per
$N_{side}$, the number of side atoms on the top layer ($N_{side}$) (equal
the number of inter-layer bonds on the edges), approaches a constant value of
$14.3$ meV as shown in Fig.4b.  However, the FM-AF transition voltage $V_c$
decreases with the system size as can be seen from Fig.4c. For the largest
system size studied, $N=1507$, we obtain $\Delta V_c=0.345$ eV, which
corresponds to an electrical field of $\sim 1$ V/nm, a value within
experimental range\cite{MLS+09}.

\begin{figure}
\epsfig{file=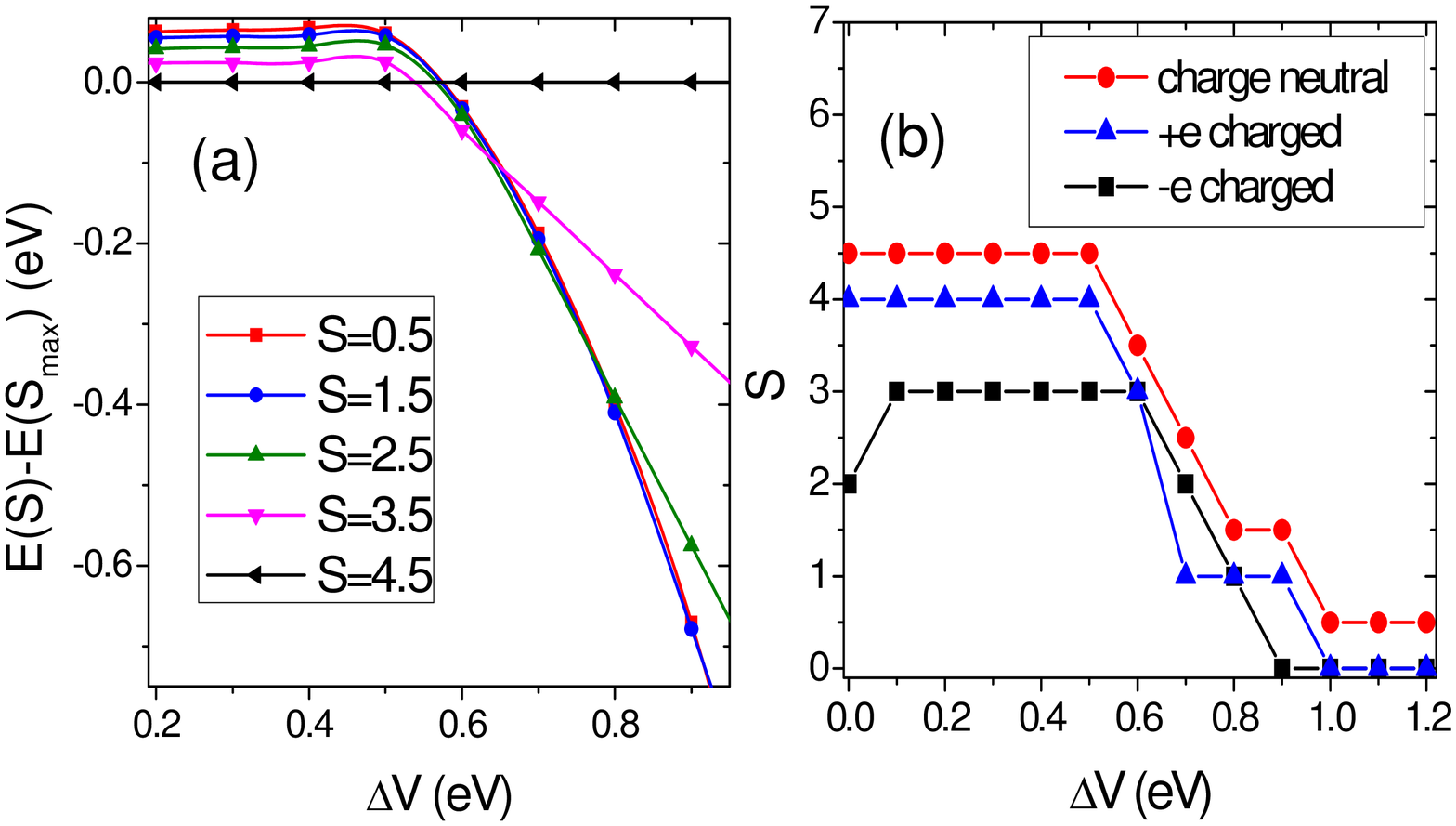,width=2.8in}
\caption{\label{fig:5} (Color online)  
Configuration interaction results. (a) Energies of states for different total
spin $S$ as a function of potential difference $\Delta V$ between the layers, with
respect to the ferromagnetic configuration $S_{max}=4.5$ for the same system as in
Fig.3. (b) Ground state total spin $S$ as a function of $\Delta V$ for the charge neutral
system, $-e$ charged system, and $+e$ charged system.}

\end{figure}

We now test the predictions of the mean-field Hubbard model by including long
range interaction and correlation effects, which is computationally feasible
for the system size studied in Fig.3. We first solve a Hartree-Fock
generalization of the Eq.2, but with empty zero-energy states\cite{GPV+09}:

\begin{eqnarray}
\nonumber
&H_{HF}&= \sum_{ij\sigma}\tau_{ij}c^\dagger_{i\sigma}c_{j\sigma} +
\sum_{i\sigma}V_{i}n_{i\sigma} + \sum_{il\sigma} \sum_{jk\sigma'} (\rho_{jk\sigma'}-\rho^{bulk}_{jk\sigma'}) \\
&\times& (\langle ij\vert V_{ee} \vert kl\rangle
-\langle ij\vert V_{ee} \vert lk\rangle\delta_{\sigma,\sigma'})
           c^\dagger_{i\sigma}c_{l\sigma} 
\label{eq:Hhf}
\end{eqnarray}

where the tight-binding term now includes the intra-layer next-nearest
neighbour hopping $t=-0.1$ eV, inter-layer next-nearest hoppings
$\gamma_3=0.3$ eV and $\gamma_4=0.15$ eV\cite{CNM+10}. The terms $\rho$ and
$\rho^{bulk}$ are quantum dot and bulk density matrices, respectively
\cite{GPV+09}. The two-body Coulomb matrix elements $\langle ij\vert V_{ee} \vert
kl\rangle$ computed using Slater $p_z$ orbitals include on-site interactions,
all scattering and exchange terms within next-nearest neighbors, and all long
range direct interactions. We have previously tested the validity of our
approach by comparing to density functional calculations and obtained good
agreement\cite{GPV+09}. After diagonalizing the Hartree-Fock Hamiltonian, we
obtain Hartree-Fock quasi-particles denoted by the creation operator
$b^\dagger_{p\sigma}$, with eigenvalues $\epsilon_p$ and eigenfunctions $\vert
p \rangle$. We can now fill the new quasi-particle zero-energy states with
electrons and solve the many-body Hamiltonian given by 

\begin{eqnarray}
H= \sum_{p\sigma}\epsilon_{p}b^\dagger_{p\sigma}b_{p\sigma}
    +\frac{1}{2}\sum_{pqrs\sigma \sigma'}\langle pq\vert V_{ee} \vert rs\rangle 
     b^\dagger_{p\sigma}b^\dagger_{q\sigma'}b_{r\sigma'}b_{s\sigma}
\label{CIhamilton}
\end{eqnarray}

Figure 5a shows the evolution of spin states for the same system studied in
Fig.3, but now obtained by diagonalizing the many-body Hamiltonian. We observe
two main differences from the mean-field Hubbard results: (i) At low $\Delta
V$, the antiferromagnetic configuration $S=1/2$ is no longer the first excited
state. Although the AFM-FM energy gap is still comparable to the Hubbard
result, other spin excitations are now closer to the ground state due to
correlation effects. (ii) The ground state spin transitions do not occur as
abruptly as in a Hubbard model. As shown in Fig.5b, the total spin of the
charge neutral system (red color online, solid line with circles) evolves
towards the minimum spin state $S=1/2$ gradually between $\Delta V=0.5-1.0$
eV. This is mainly due to long range interactions. As the electrons are
transferred one by one from the top layer into the bottom layer, they leave a
positively charged hole behind which makes it harder to transfer more
electrons. We note that this behavior of gradual spin transition is also
obtained within a mean-field Hubbard model with long range interactions
included (not shown). Finally, in Fig.5b, we also study  the effect of
charging the BQD system. The asymmetry between magnetic moment of $+e$ and
$-e$ charged systems reflects the correlation induced spin depolarization
process that occurs in single layer TGQDs as discussed in our previous
works\cite{GPV+09}.

In summary,  we demonstrate that the graphene bilayer triangular quantum dots
exhibit a shell of degenerate  states at the Fermi level. At half filling, the
shell is maximally spin polarized.  By the application of a vertical electric
field the total spin of the bilayer structure  can be turned off or reduced to
a single localized spin, a qubit isolated from contacts and free from
interaction with nuclear spins. This opens new possibilities in the area of
spintronics and quantum information processing.

{\it Acknowledgment}. The authors thank NRC-CNRS CRP, Canadian Institute for
Advanced Research, Institute for Microstructural Sciences, and QuantumWorks
for support. 


\vspace*{-0.22in}


\begin{thebibliography}{}
\expandafter\ifx\csname natexlab\endcsname\relax\def\natexlab#1{#1}\fi
\expandafter\ifx\csname bibnamefont\endcsname\relax
  \def\bibnamefont#1{#1}\fi
\expandafter\ifx\csname bibfnamefont\endcsname\relax
  \def\bibfnamefont#1{#1}\fi
\expandafter\ifx\csname citenamefont\endcsname\relax
  \def\citenamefont#1{#1}\fi
\expandafter\ifx\csname url\endcsname\relax
  \def\url#1{\texttt{#1}}\fi
\expandafter\ifx\csname urlprefix\endcsname\relax\def\urlprefix{URL }\fi
\providecommand{\bibinfo}[2]{#2}
\providecommand{\eprint}[2][]{\url{#2}}


\bibitem[{\citenamefont{Wallace}(1947)}]{Wallace}
\bibinfo{author}{\bibfnamefont{P.~R.}~\bibnamefont{Wallace}},
  \bibinfo{journal}{Phys. Rev.} \textbf{\bibinfo{volume}{71}}
  , \bibinfo{pages}{622} (\bibinfo{year}{1947}).

\bibitem[{\citenamefont{Novoselov et~al.}(2004)\citenamefont{Novoselov, Geim,
  Morozov, Jiang, Zhang, Dubonos, Grigorieva, and Firsov}}]{NGM+04}
\bibinfo{author}{\bibfnamefont{K.~S.} \bibnamefont{Novoselov}},
  \bibinfo{author}{\bibfnamefont{A.~K.} \bibnamefont{Geim}},
  \bibinfo{author}{\bibfnamefont{S.~V.} \bibnamefont{Morozov}},
  \bibinfo{author}{\bibfnamefont{D.}~\bibnamefont{Jiang}},
  \bibinfo{author}{\bibfnamefont{Y.}~\bibnamefont{Zhang}},
  \bibinfo{author}{\bibfnamefont{S.~V.} \bibnamefont{Dubonos}},
  \bibinfo{author}{\bibfnamefont{I.~V.} \bibnamefont{Grigorieva}},
  \bibnamefont{and} \bibinfo{author}{\bibfnamefont{A.~A.} \bibnamefont{Firsov}}, 
\bibinfo{journal}{Science}
  \textbf{\bibinfo{volume}{306}}, \bibinfo{pages}{666} (\bibinfo{year}{2004}).

\bibitem[{\citenamefont{Novoselov et~al.}(2005)\citenamefont{Novoselov, Geim,
  Morozov, Jiang, Katsnelson, Grigorieva, Dubonos, and Firsov}}]{NGM+05}
\bibinfo{author}{\bibfnamefont{K.~S.} \bibnamefont{Novoselov}},
  \bibinfo{author}{\bibfnamefont{A.~K.} \bibnamefont{Geim}},
  \bibinfo{author}{\bibfnamefont{S.~V.} \bibnamefont{Morozov}},
  \bibinfo{author}{\bibfnamefont{D.}~\bibnamefont{Jiang}},
  \bibinfo{author}{\bibfnamefont{M.~I.} \bibnamefont{Katsnelson}},
  \bibinfo{author}{\bibfnamefont{I.~V.} \bibnamefont{Grigorieva}},
  \bibinfo{author}{\bibfnamefont{S.~V.} \bibnamefont{Dubonos}},
  \bibnamefont{and} \bibinfo{author}{\bibfnamefont{A.~A.} \bibnamefont{Firsov}}, 
\bibinfo{journal}{Nature}
  \textbf{\bibinfo{volume}{438}}, \bibinfo{pages}{197} (\bibinfo{year}{2005}).

\bibitem[{\citenamefont{Zhang et~al.}(2005)\citenamefont{Zhang, Tan, Stormer,
  and Kim}}]{ZTS+05}
\bibinfo{author}{\bibfnamefont{Y.~B.} \bibnamefont{Zhang}},
  \bibinfo{author}{\bibfnamefont{Y.~W.} \bibnamefont{Tan}},
  \bibinfo{author}{\bibfnamefont{H.~L.} \bibnamefont{Stormer}},
  \bibnamefont{and} \bibinfo{author}{\bibfnamefont{P.}~\bibnamefont{Kim}},
  \bibinfo{journal}{Nature} \textbf{\bibinfo{volume}{438}},
  \bibinfo{pages}{201} (\bibinfo{year}{2005}).

\bibitem[{\citenamefont{Zhou et~al.}(2006)\citenamefont{Zhou, Gweon, Graf,
  Fedorov, Spataru, Diehl, Kopelevich, Lee, Louie, and Lanzara}}]{ZGG+06}
\bibinfo{author}{\bibfnamefont{S.~Y.} \bibnamefont{Zhou}},
  \bibinfo{author}{\bibfnamefont{G.~H.} \bibnamefont{Gweon}},
  \bibinfo{author}{\bibfnamefont{J.}~\bibnamefont{Graf}},
  \bibinfo{author}{\bibfnamefont{A.~V.} \bibnamefont{Fedorov}},
  \bibinfo{author}{\bibfnamefont{C.~D.} \bibnamefont{Spataru}},
  \bibinfo{author}{\bibfnamefont{R.~D.} \bibnamefont{Diehl}},
  \bibinfo{author}{\bibfnamefont{Y.}~\bibnamefont{Kopelevich}},
  \bibinfo{author}{\bibfnamefont{D.~H.} \bibnamefont{Lee}},
  \bibinfo{author}{\bibfnamefont{S.~G.} \bibnamefont{Louie}}, \bibnamefont{and}
  \bibinfo{author}{\bibfnamefont{A.}~\bibnamefont{Lanzara}},
  \bibinfo{journal}{Nature Phys.} \textbf{\bibinfo{volume}{2}},
  \bibinfo{pages}{595} (\bibinfo{year}{2006}). 

\bibitem[{\citenamefont{Neto et~al.}(2009)\citenamefont{Neto, Guinea, Peres,
  Novoselov, and Geim}}]{NGP+09}
\bibinfo{author}{\bibfnamefont{A.~H.~C.} \bibnamefont{Neto}},
  \bibinfo{author}{\bibfnamefont{F.}~\bibnamefont{Guinea}},
  \bibinfo{author}{\bibfnamefont{N.~M.~R.} \bibnamefont{Peres}},
  \bibinfo{author}{\bibfnamefont{K.~S.} \bibnamefont{Novoselov}},
  \bibnamefont{and} \bibinfo{author}{\bibfnamefont{A.~K.} \bibnamefont{Geim}},
  \bibinfo{journal}{Rev. of Mod. Phys.} \textbf{\bibinfo{volume}{81}},
  \bibinfo{pages}{109} (\bibinfo{year}{2009}). 



\bibitem[{\citenamefont{Ponomarenko et~al.}(2008)\citenamefont{Ponomarenko,
  Schedin, Katsnelson, Yang, Hill, Novoselov, and Geim}}]{PSK+08}
\bibinfo{author}{\bibfnamefont{L.~A.} \bibnamefont{Ponomarenko}},
  \bibinfo{author}{\bibfnamefont{F.}~\bibnamefont{Schedin}},
  \bibinfo{author}{\bibfnamefont{M.~I.} \bibnamefont{Katsnelson}},
  \bibinfo{author}{\bibfnamefont{R.}~\bibnamefont{Yang}},
  \bibinfo{author}{\bibfnamefont{E.~W.} \bibnamefont{Hill}},
  \bibinfo{author}{\bibfnamefont{K.~S.} \bibnamefont{Novoselov}},
  \bibnamefont{and} \bibinfo{author}{\bibfnamefont{A.~K.} \bibnamefont{Geim}},
  \bibinfo{journal}{Science} \textbf{\bibinfo{volume}{320}},
  \bibinfo{pages}{356} (\bibinfo{year}{2008}).

\bibitem{LYG+11}
Jiong Lu, Pei Shan Emmeline Yeo, Chee Kwan Gan, Ping Wu and Kian Ping Loh,
{ Nature Nanotechnology} {\bf 6}, 247–252 (2011).

\bibitem{GFS+10}
J. Guttinger, T. Frey, C. Stampfer, T. Ihn, and K. Ensslin,  
{ Phys. Rev. Lett.} {\bf 105}, 116801 (2010).

\bibitem[{\citenamefont{Wunsch et~al.}(2008{\natexlab{b}})\citenamefont{Wunsch,
  Stauber, and Guinea}}]{WSG08}
\bibinfo{author}{\bibfnamefont{B.}~\bibnamefont{Wunsch}},
  \bibinfo{author}{\bibfnamefont{T.}~\bibnamefont{Stauber}}, \bibnamefont{and}
  \bibinfo{author}{\bibfnamefont{F.}~\bibnamefont{Guinea}},
  \bibinfo{journal}{Phys. Rev. B} \textbf{\bibinfo{volume}{77}},
  \bibinfo{pages}{035316} (\bibinfo{year}{2008}{\natexlab{b}}).

\bibitem[{\citenamefont{Akola et~al.}(2008)\citenamefont{Akola, Heiskanen, and
  Manninen}}]{AHM08}
\bibinfo{author}{\bibfnamefont{J.}~\bibnamefont{Akola}},
  \bibinfo{author}{\bibfnamefont{H.~P.} \bibnamefont{Heiskanen}},
  \bibnamefont{and} \bibinfo{author}{\bibfnamefont{M.}~\bibnamefont{Manninen}},
  \bibinfo{journal}{Phys. Rev. B} \textbf{\bibinfo{volume}{77}},
  \bibinfo{pages}{193410} (\bibinfo{year}{2008}).

\bibitem[{\citenamefont{Ezawa}(2010)}]{Eza10}
\bibinfo{author}{\bibfnamefont{M.}~\bibnamefont{Ezawa}},
  \bibinfo{journal}{Phys. Rev. B} \textbf{\bibinfo{volume}{81}},
  \bibinfo{pages}{201402} (\bibinfo{year}{2010}). 


\bibitem[{\citenamefont{G\"u\c{c}l\"u et~al.}(2009)\citenamefont{G\"u\c{c}l\"u, Potasz, Voznyy,
  Korkusinski, and Hawrylak}}]{GPV+09}
\bibinfo{author}{\bibfnamefont{A.~D.} \bibnamefont{G\"u\c{c}l\"u}},
  \bibinfo{author}{\bibfnamefont{P.}~\bibnamefont{Potasz}},
  \bibinfo{author}{\bibfnamefont{O.}~\bibnamefont{Voznyy}},
  \bibinfo{author}{\bibfnamefont{M.}~\bibnamefont{Korkusinski}},
  \bibnamefont{and} \bibinfo{author}{\bibfnamefont{P.}~\bibnamefont{Hawrylak}},
  \bibinfo{journal}{Phys. Rev. Lett.} \textbf{\bibinfo{volume}{103}},
  \bibinfo{pages}{246805}  (\bibinfo{year}{2009}).

\bibitem{GPH10}
A. D. G\"u\c{c}l\"u, P. Potasz, and P. Hawrylak, 
{ Phys. Rev. B} {\bf 82}, 155445 (2010).

\bibitem[{\citenamefont{Potasz et~al.}(2010)\citenamefont{Potasz, G\"u\c{c}l\"u, and
  Hawrylak}}]{PGH10}
\bibinfo{author}{\bibfnamefont{P.}~\bibnamefont{Potasz}},
  \bibinfo{author}{\bibfnamefont{A.~D.} \bibnamefont{G\"u\c{c}l\"u}}, \bibnamefont{and}
  \bibinfo{author}{\bibfnamefont{P.}~\bibnamefont{Hawrylak}},
  \bibinfo{journal}{Phys. Rev. B} \textbf{\bibinfo{volume}{81}},
  \bibinfo{pages}{033403}  (\bibinfo{year}{2010}).

\bibitem{VGP+11}
O. Voznyy, A.D. G\"u\c{c}l\"u, P. Potasz , P. Hawrylak, 
{ Phys.Rev.B.} (in press).

\bibitem[{\citenamefont{Fernandez-Rossier and Palacios}(2007)}]{FP07}
\bibinfo{author}{\bibfnamefont{J.}~\bibnamefont{Fernandez-Rossier}}
  \bibnamefont{and} \bibinfo{author}{\bibfnamefont{J.~J.}
  \bibnamefont{Palacios}}, \bibinfo{journal}{Phys. Rev. Lett.}
  \textbf{\bibinfo{volume}{99}}, 
  \bibinfo{pages}{177204} (\bibinfo{year}{2007}).

\bibitem[{\citenamefont{Wang et~al.}(2008)\citenamefont{Wang, Meng, and
  Kaxiras}}]{WMK08}
\bibinfo{author}{\bibfnamefont{W.~L.} \bibnamefont{Wang}},
  \bibinfo{author}{\bibfnamefont{S.}~\bibnamefont{Meng}}, \bibnamefont{and}
  \bibinfo{author}{\bibfnamefont{E.}~\bibnamefont{Kaxiras}},
  \bibinfo{journal}{Nano Letters} \textbf{\bibinfo{volume}{8}},
  \bibinfo{pages}{241} (\bibinfo{year}{2008}).



\bibitem{ENM+09}
D. C. Elias 
, R. R. Nair, T. M. G. Mohiuddin, S. V. Morozov, P. Blake, M. P. Halsall,
A. C. Ferrari, D. W. Boukhvalov, M. I. Katsnelson, A. K. Geim and K. S. Novoselov,
{ Science} {\bf 323} 610-613 (2009).


\bibitem{MAF07}
E. McCann 
, D. S. L. Abergel, V. I. Fal’ko,
{ Solid St. Comm.} {\bf 143}, 110 (2007).

\bibitem{OBS+06}
T. Ohta 
, A. Bostwick, T. Seyller, K. Horn and E. Rotenberg,  
{ Science} {\bf 313}, 951 (2006).

\bibitem{CNM+07}
Eduardo V. Castro 
, K. S. Novoselov, S. V. Morozov, N. M. R. Peres,
J. M. B. Lopes dos Santos, Johan Nilsson, F. Guinea, A. K. Geim, and
A. H. Castro Neto,
{ Phys. Rev. Lett.} {\bf 99}, 216802 (2007).

\bibitem{OHL+07}
J. B. Oostinga 
, Hubert B. Heersche, Xinglan Liu, Alberto F. Morpurgo and
Lieven M. K. Vandersypen,
{ Nat. Mater.} {\bf 7}, 151 (2007).

\bibitem{MLS+09}
K. F. Mak 
, C. H. Lui, J. Shan, and T. F. Heinz,
{ Phys. Rev. Lett.} {\bf 102}, 256405 (2009).

\bibitem{ZTG+09}
Y. Zhang 
T. T. Tang, C. Girit, Z. Hao, M. A. Martin, A. Zettl,
M. F. Crommie, Y. R. Shen, F. Wang,
{ Nature} {\bf 459}, 820 (2009).

\bibitem{WAF+10}
R. T. Weitz 
, M. T. Allen, B. E. Feldman, J. Martin and A. Yacoby,
{Science} {\bf 330}, 812 (2010).

\bibitem{CNM+10}
Eduardo V. Castro 
K. S. Novoselov, S. V. Morozov, N. M. R. Peres,
J. M. B. Lopes dos Santos, Johan Nilsson, F. Guinea, A.K. Geim and
A.H. Castro Neto,  
{ J. Phys.:Condens. Matter}, {\bf 22}, 175503 (2010).

\bibitem{ZLB+08}
L.M. Zhang 
, Z. Q. Li, Dimitri N. Basov, M.M. Fogler, Zhao Hao, Michael C. Martin,
{ Phys. Rev. B}, {\bf 78}, 235408 (2008).


\bibitem{PVP07}
J. Milton Pereira, P. Vasilopoulos, and F. M. Peeters,
Nano Lett., {\bf 7}, 946 (2007).

\bibitem{CPS+08}
Eduardo V. Castro 
, N. M. R. Peres, J. M. B. Lopes dos Santos, A. H. Castro
Neto, and F. Guinea,
Phys. Rev. Lett. {\bf 100}, 026802 (2008).

\bibitem{SMM+08}
Bhagawan Sahu 
, Hongki Min, A. H. MacDonald, and Sanjay K. Banerjee,
Phys. Rev. B {\bf 78}, 045404 (2008).

\bibitem{WCZ09}
A. R. Wright, J. C. Cao and C. Zhang,
Phys. Rev. Lett. {\bf 103}, 207401 (2009).




\bibitem{OCM+00}
H. Ohno  
D. Chiba, F. Matsukura, T. Omiya, E. Abe, T. Dietl1, Y. Ohno and K. Ohtani,
{ Nature} {\bf 408}, 944 (2000).

\bibitem{RTB07}
A. Rycerz, J. Tworzydo and C. W. J. Beenakker, 
{ Nature Physics} {\bf 3}, 172 (2007).

\bibitem{SCL06}
Y. W. Son, M. L. Cohen, and S. G. Louie, 
{ Nature} {\bf 444}, 347 (2006)

\bibitem{TBL+07}
Björn Trauzettel 
Denis V. Bulaev, Daniel Loss and Guido Burkard, 
{ Nature Physics} {\bf 3}, 192 (2007). 

\bibitem{SSC10}
H. Sahin, R. T. Senger and S. Ciraci,
{ J. Appl. Phys.} {\bf 108}, 074301 (2010).

\bibitem{AKK10}
L. A. Agapito1, N. Kioussis1, and E. Kaxiras,
{ Phys. Rev. B} {\bf 82}, 201411(R) (2010).


\bibitem{Lieb89}
E. H. Lieb,
{ Phys. Rev. Lett.} {\bf 62}, 1201 (1989).





\end{thebibliography}

\end{document}